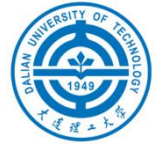

# Mining Meta-indicators of University Ranking: A Machine Learning Approach Based on SHAP

Shudong YANG[1], Miaomiao LIU[1]

(1. Dalian University of Technology, Dalian, China, 116024)

**Abstract**: [**Background**] University evaluation and ranking is an extremely complex activity. Major universities are struggling because of increasingly complex indicator systems of world university rankings. [**Question**] So can we find the meta-indicators of the index system by simplifying the complexity? [**Approach**] This research discovered three meta-indicators based on interpretable machine learning. [**Results**] The first one is time, to be friends with time, and believe in the power of time, and accumulate historical deposits; the second one is space, to be friends with city, and grow together by co-develop; the third one is relationships, to be friends with alumni, and strive for more alumni donations without ceiling.

**Keywords**：meta-indicators; university ranking; SHAP; interpretable machine learning

## 1. Introduction

The ultimate goal of any type of evaluation is not the evaluation itself, but the event outside of the evaluation. There are many elements to form an evaluation organism. The establishment of a university ranking evaluation index system must be built from the consensus of practice and theory, and any set of evaluation indexes is set for evaluation of specific research objects. The traditional index system of university rankings includes talent training, scientific research, social services, etc. The index system constructed by the major rankings may have certain theoretical support or certain practical feedback.

Taking ARWU as an example, there are three levels of indicators. The first-level indicators include school level, discipline level, resources, teacher scale and structure, talent training, scientific research, social services, academic talents, major projects and achievements, and international competitiveness total 10 items, and then subdivided into 35 second-level indicators, hundreds of third-level indicators, and hundreds of evaluation variables[1].

---

However, there are some problems in the existing index system. First, regional location is crucial to the development of universities. For example, universities in fast-developing cities such as Shenzhen in China have a greater talent pool advantage than other universities. There is no variable that characterizes space in the existing index system. Second, there is no time-related or historical background variables. Third, there is col-linearity between indicators, such as funding indicators and many other indicators are related, many indicators are ultimately funding issues. Fourth, the indicator system is becoming more and more complex, which may cover up the real key influencing factors. Therefore, this research aims to mine which meta-indicators other than existing indicators are useful for explaining university rankings.

## 2. Data Processing

This research takes the top 100 universities in China ranked by ARWU as the research object. The dependent variable is the total score of ARWU for a particular university in 2021. The independent variables take into account the impact of the university's history, urban economic development, national and regional policy systems, university education funding, etc., and summarize the four major dimensional variables that affect university rankings. There are four types of independent variables and a total of 11 variables as follow table:

**Table 1: variables that may affect university ranking**

| Dimension | Variable |
|---|---|
| Funding input | • Taking into account the lag in performance, the education funding data is taken from the budgets of major universities in the previous year (2020). |
| Length of university establishment | • The variable of length of university establishment represents the historical background, the data is taken from the introduction of Baidu Encyclopedia. |
| Policy | • The growth rate of undergraduate enrollment in the past five years. Enrollment growth is a function of policy in China. |
| City development competitiveness | • Per capita GDP of the province where the university is located;<br>• The GDP growth trend of the province where the university is located in the past 10 years;<br>• The annual average temperature;<br>• Whether the province where the university is located is coastal;<br>• The average housing price in the location of the university;<br>• The average price of a three-star hotel where the university is located;<br>• The box office of the province where the university is located;<br>• The number of cars owned by one hundred people in the province where the university is located; |

Mining Meta-indicators of University Ranking: A Machine Learning Approach Based on SHAP

There are 88 valid samples after excluding universities with missing funding data. An 88*12 matrix composed of continuous variables in closed intervals of [0, 1] is obtained by data preprocessing. In order to explain the importance of each variable of the machine learning model, there are two main methods: feature attribution methods, such as SHAP[2][3], and counterfactual interpretation methods, such as DiCE. The two methods may be inconsistent in the ranking of feature importance. This research uses the former to explain. Under the constraints of MSE less than 0.1 and R^2 greater than 0.8, the trained model is obtained by LightGBM machine learning, and the SHAP method is used to explain the contribution of each variable to the overall model, as shown in the following figure:

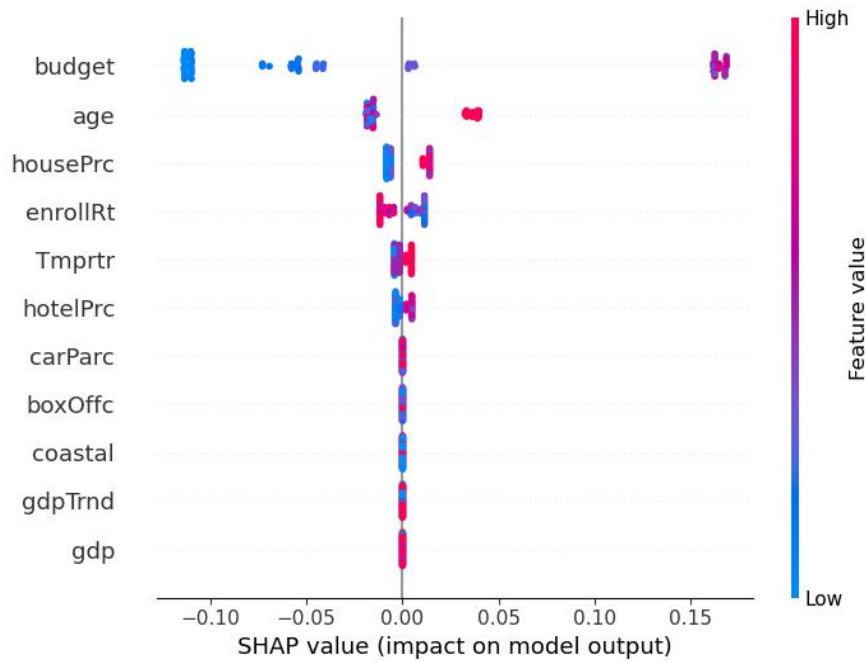

**Figure 1: the SHAP value graph of the potential meta-indicator of university ranking**

The upper independent variable in the figure is more explanatory to the model as a whole. Each scatter point represents a sample (Top 100 Universities in China). The color of the scattered points represents the size of the independent variable (**pink** is large, **light blue** is small). The horizontal axis represents the degree of influence of the independent variable on the dependent variable. The same independent variable has different degrees of influence on different universities, which is also in line with the actual situation. The vertical axis is meaningless, but when the SHAP values of multiple samples are repeated, it will jitter up and down, visually forming a vertical distribution. The specific explanation is as follows.

# 3. Meta-indicators Mining

Education funding (variable: *budget*) is positively correlated with the dependent variable. With the development of the economy, the investment in education funds is gradually increasing. On the one hand, the education funds in the economically developed areas are relatively large, while in the economically backward areas, the education funds are relatively weak, which intensifies the inter-regional financial investment. Unbalanced development, education funding investment is bound to affect the long-term development of regional universities. On the other hand, the educational expenditure of "first-class universities" and "first-class disciplines" universities is significantly higher than that of other universities or disciplines. For "high investment and high return", it will inevitably affect the overall ranking of universities.

---

[2] C Molnar. Interpretable Machine Learning: A Guide for Making Black Box Models Explainable[M]. lulu.com, 2020.
[3] S Masís. Interpretable Machine Learning with Python: Learn to build interpretable high-performance models with hands-on real-world examples[M]. Packt Publishing, 2021.

The length of university establishment (variable: *age*), the top universities are almost with a long history, and vice versa. The history of establishment and development of a university is the epitome of the history of higher education in a country or region. Generally speaking, the longer the establishment of a university, the deeper the historical and cultural accumulation, the more complete the resources and infrastructure the university has, the richer the operation experience, and the stronger the overall ability of the university, which affects the overall ranking of the university. But history itself does not represent the status of a university. There are many universities with a long history that are not world-class universities, and there are also a few successful examples of universities that have only been established for 50 or 60 years, such as Swiss federal Institute of Technology in Lausanne (1969), Nanyang Technological University (1955), The Hong Kong University of Science and Technology, China (1986), Korea Advanced Institute of Science and Technology (1971), City University of Hong Kong, China (1984), etc.

The average house price (variable: *housePrc*) at the location of the university is positively correlated with the dependent variable. The housing price is one of the manifestations of the development level of a city. The higher the housing price, the more developed the regional economy. The regional economy of a city and the development of higher education interact and promote each other.

For other factors, the 4th place is the growth rate of undergraduate enrollment in the past five years (variable: *enrollRt*), which is negatively correlated with the dependent variable. In particular, in China, the scale of enrollment is not a function of the market, but a function of policy. The 5th place is the annual average temperature of the city where the university is located (variable: *Tmprtr*), which represents the North-South difference in China; The 6th place is the average price of a three-star hotel where the university is located (variable: *hotelPrc*), which represents the level of development and vitality of the city. The other remaining variables are less explanatory.

## 4. Conclusion

In summary, for university managers, rather than struggling to target more and more complex evaluation indicators, **it is better to go back to the roots and accumulate true power from meta-indicators**: The first one is time, **to be friends with time**, and believe in the power of time, and accumulate historical deposits; the second one is space, **to be friends with city**, and grow together by co-develop; the third one is relationships, **to be friends with alumni**, and strive for more alumni donations without ceiling.